\documentstyle[12pt]{article}
\def\be{\begin{equation}}
\def\ee{\end{equation}}
\def\ba{\begin{eqnarray}}
\def\ea{\end{eqnarray}}

\def\bq{\begin{quote}}
\def\eq{\end{quote}}
\def\PL{{ \it Phys. Lett.} }
\def\PRL{{\it Phys. Rev. Lett.} }
\def\NP{{\it Nucl. Phys.} }
\def\PR{{\it Phys. Rev.} }

\newcommand{\beq}{\begin{equation}}
\newcommand{\eeq}{\end{equation}}
\newcommand{\beqa}{\begin{eqnarray}}
\newcommand{\eeqa}{\end{eqnarray}}

\begin{document}
\thispagestyle{empty}
\begin{flushright}
SU-ITP-99/25\\ hep-th/9905210\\ May 1999
\end{flushright}
\vspace*{1cm}
\begin{center}
{\Large \bf Bent Domain Walls as Braneworlds}\\
\vspace*{1.5cm}
Nemanja Kaloper\footnote{email: kaloper@epic.stanford.edu}\\
\vspace*{0.2cm}
{\it Department of Physics}\\
{\it Stanford University}\\
{\it Stanford, CA 94305-4060}\\
\vspace{1.5cm}
ABSTRACT
\end{center}
We consider domain walls embedded in curved backgrounds
as an approximation for braneworld scenarios. We give a large
class of new exact solutions, exhausting the possibilities for
describing one and two walls for the cases where the curvature of both
the bulk and the wall is locally constant.
In the case of two walls, we find 
solutions where each wall has positive tension. 
An interesting property of
these solutions is that the curvature of the walls can be
much smaller than the tension, leading to a significant
cancellation of the effective cosmological constant, which however
is still much larger than the observational limits.
We further discuss some aspects of inflation in models based
on wall solutions. 
\vfill
\setcounter{page}{0}
\setcounter{footnote}{0}
\newpage

\section{Introduction}

There has been considerable interest recently in higher dimensional
theories with internal dimensions which are large compared to the
fundamental Planck length of the theory \cite{savas}-\cite{ahdkmr}.
If the fundamental theory is higher-dimensional,
the higher-dimensional fundamental Planck scale may be many
orders of magnitude smaller than the Planck scale in a world
of only four dimensions ($4D$).
The effective $4D$ Planck mass is given by
$M^2_4 \sim \Bigl(\frac{r_0}{l_{D+4}}\Bigr)^D M^2_{D+4}$
where $r_0$ is the size of the compact dimensions
\cite{savas}.
If these dimensions are much larger than the
fundamental Planck length,
$\frac{r_0}{l_{D+4}} >>1$, the reduced Planck mass may be
exponentially larger than the fundamental Planck mass.
Setting the fundamental Planck mass to be equal to the
unification scale of the theory, 
one may then understand the mass hierarchy
as a consequence of stabilizing the internal space
at a radius which is exponentially large compared to the
fundamental scale. This scale has been elucidated in \cite{ant}
as emerging from supersymmetry breaking problem in superstring theory.

In most of the models considered to date,
the $4D$ world is a domain wall, or
$3$-brane, living in a higher-dimensional theory with compact
internal dimensions. The metric was independent of
the internal coordinates.
Similar proposals have been made before \cite{rus,ds},
but recent developments in string and
$M$-theory, which may explain why matter
degrees of freedom are stuck to the wall \cite{hw,string} gave new
credibility to such ideas.

It has been pointed recently, however, that a different realization
of the higher-dimensional resolution of mass hierarchy may
be found if the metric depends on internal coordinates \cite{rs}.
Specifically, the hierarchy emerges because the
scales change throughout the internal dimensions, and
the particle masses are literally a function of the location of our
$3$-brane in the internal space. In a very interesting work, Randall
and Sundrum \cite{rs} have shown that even if there is only one extra
dimension, and even if it remains very small, it may still be
possible to obtain large mass hierarchies if the extra
dimension is strongly curved. In their approach, this
was provided by a large negative cosmological constant
in the extra dimensions, which makes the metric a very
sensitive function of the internal coordinates. Still, this
bulk vacuum energy is cancelled on 
the brane by the brane cosmological constant,
and so it would be cosmologically neutralized inside our world.

In this work we will explore the solutions where the
bulk cosmological constant is not exactly cancelled
by the brane one. Our motivation for this is twofold. First,
we are interested in determining the circumstances under which
it may be possible to have the
brane universes which may be rapidly expanding, and hence
lead to successful inflationary cosmology. Existence
of such scenarios is very important in models with low
compactification scale. In addition to the usual
cosmological problems of the $4D$ universe, there one has to
explain the emptiness of large extra dimensions, which could
contain far more energy than the brane. Since this
would contradict the observations, such cosmological
obstacles need to be answered dynamically, and inflation 
is the best candidate \cite{savas,kl,ahdkmr}.
In the models with sub-millimeter internal dimension
this problem is ably handled by inflation before
stabilization of extra dimensions \cite{ahdkmr}.
Second, it may be possible to find
compactifications which are different from the example
given in \cite{rs}, but where nevertheless
the mass hierarchy
emerges because of the strongly curved internal dimensions.
Indeed, we will present new solutions
describing parallel branes.
Each brane will have a small net cosmological
constant. In some cases the net cosmological
constant will be negative, which may be less attractive, but
this possibility is not ruled out experimentally.
The effective cosmological constant on each brane
may be many orders of magnitude smaller than the
natural scale due to the cancellation between the bulk
cosmological constant and the brane tension. However, with the
phenomenologically reasonable
parameters needed to generate hierarchy,
we find that the cancellation still falls short
of producing an effective cosmological constant
in the range of $10^{-120} M^4_{Pl}$. Nevertheless, the mere
fact that it may be possible to get a small
cosmological constant from large ones
in this way deserves notice. 
There may exist other solutions which are
realized in a similar spirit, that yield
an even smaller nonzero cosmological constant on the wall.

Throughout the article we will
also discuss some issues of relevance for early cosmology
and inflation on branes in curved backgrounds. Rather
interestingly, many inflating branes get surrounded by
event horizons in the bulk during inflation. As inflation
nears the end, these horizons move slowly away from the brane.
If the inflationary periods are long, the dynamics of the
bulk event horizon may help solve
the bulk horizon problem without significant fine tuning,
since perturbations inside of the horizons get stretched
and pushed far away (to the hidden wall?).
Hence, such solutions may lead to economical
higher-dimensional inflationary scenarios.

The paper is organized as follows. We begin by deriving
the solutions corresponding to one and two branes
in bulks with constant curvature. We then proceed to discuss
the properties of brane spacetimes, discussing the generalization
of the usual Friedmann equation in cosmology and the role
of the bulk event horizons. In Section IV we consider the
features of the new solution with two parallel branes, noting that
the effective cosmological constant on each brane
could be much smaller than the
natural scale, although still too large to fit the observations.
Finally, we close with the Conclusion. 
The Appendix contains details
of the derivation of solutions.

\section{Solutions}

We begin by considering a $3D$ domain wall, or a $3$-brane,
in a curved $5D$ bulk, which 
couples gravitationally to the $5D$ bulk theory.
The action specifying the dynamics of the brane-bulk system is
\be
S = \int_M d^5x \sqrt{g} 
\Bigl\{\frac{R}{2\kappa^2_5} - {\cal L}_{bulk} \Bigr\}
- \int_{\partial M} d^4x \sqrt{g} {\cal L}_{brane}
\label{actdef}
\ee
Our conventions are $R^{\mu}{}_{\nu\lambda\rho}
= \partial_\lambda \Gamma^\mu_{\nu\rho}
- \partial_\rho \Gamma^\mu_{\nu\lambda} + ...$, and
$\eta_{\mu\nu} = {\rm diag}(-1,1,1,1,1)$, which are common
in low energy string and M-theory.
Further, we use $\kappa_5^2 = \frac{8\pi}{M^3_*}$, where
$M_*$ is the $5D$ Planck scale, roughly equal to the
unification scale.
The subscripts $M$ and $\partial M$ in the action
stand for the $5D$ bulk and the $4D$ brane which
from the standpoint of the bulk may be viewed as
its boundary.
We note that $\sqrt{g}$ in the 5D bulk and 4D boundaries
are different, being
related by $\sqrt{g}(boundary) = \sqrt{g}(bulk)/\sqrt{g_{55}}$.
This affects the equations of motion, ensuring that the
projection of the brane's contribution to the bulk stress-energy
tensor vanishes in the directions transverse to the brane.

To obtain the equations of motion, we can now vary the action
(\ref{actdef}) in the usual fashion. However, for the purposes
of our investigation we can simplify
our considerations without sacrificing physical contents.
We are interested in determining the brane geometry
at large distances when the bulk is curved, both during
inflation on the brane(s) and in the limit when the expansion
is very slow. Thus it is sufficient to consider
solutions where the energy density on the brane is
nearly constant, and so approximate
the brane contribution to the stress-energy tensor
by the brane
tension. Further, recalling the properties
of the scalar-tensor theories
of gravity, which approximate the theory given by (\ref{actdef})
in the infrared limit, when the bulk curvature
is changing rapidly, generically
the brane will neither inflate nor
settle down to a highly homogeneous and isotropic state.
Hence, there should be epochs when we can model the
sources of curvature in the bulk by bulk cosmological terms.
Therefore, the effective action which we consider is
\be
S = \int_M d^5x \sqrt{g} \Bigl\{\frac{R}{2\kappa^2_5}
+ \Lambda \Bigr\} - \int_{\partial M} d^4x \sqrt{g} \sigma
\label{action}
\ee
where we have taken the negative cosmological term in the bulk.
Our motivation for this sign choice is that
such models can be naturally obtained by compactifying
supergravity theories in $10D$ by using the Freund-Rubin
ansatz \cite{fr}, for example, compactifying the
IIB string theory on $AdS_5 \times S^5$ \cite{iib}.
Other examples of domain walls in supergravity have
been considered in \cite{sugrwalls,cvetic}.
A possibility that the world may look like a domain wall
in the context of Ho\v rava-Witten theory \cite{hw}
has been pursued in \cite{ovst}, and possibilities for
inflation in this model were studied in \cite{low}.
We have in mind a particular application of the solutions
to the recent scenario proposed by Randall and Sundrum \cite{rs},
and their solutions indeed emerge as the future asymptotics of
some of the solutions presented below.
However, all solutions will depend analytically on $\Lambda$,
allowing one to continue $\Lambda$
to the range of negative values, corresponding to
a positive bulk cosmological constant.
The parameter $\sigma$ is the 
brane tension, which we will also allow to be
both positive and negative (with $\sigma > 0$ corresponding
to positive brane tension).

The equations of motion which follow from (\ref{action}) are
\be
R^{\mu}{}_{\nu} - \frac12 \delta^{\mu}{}_{\nu} R = -\kappa_5^2
{\frac{\sqrt{g_{brane}}}{\sqrt{g}}}
\sigma \delta(w) {\rm diag}(1,1,1,1,0) + \kappa_5^2
\Lambda  \delta^{\mu}{}_{\nu}
\label{ees}
\ee
where the stress-energy tensor is a combination of wall
and bulk terms. In general the ratio ${\frac{\sqrt{g_{brane}}}{\sqrt{g}}}$
cannot be gauged away.
To find the specific form of the equations
of motion, we need to write down the metric ansatz which
will reflect the symmetries on the $3$-brane. Here we
can use some known solutions as a guideline.
In the limit $\Lambda = 0$, $\sigma > 0$,
the solution must correspond
to the inflating $3$-brane in $5D$, found in \cite{kl}; it is
given by
$ds^2 = (1-\frac{4\pi \sigma}{3M^3_5}|w|)^2(-dt^2 +
\exp(8\pi \sigma t/3M^3_5) d\vec x^2) + dw^2$,
and corresponds to a $3D$ domain wall undergoing inflation,
which is shrouded by a Rindler horizon in the bulk.
In the limit $\Lambda = \kappa^2_5 \sigma^2/6$, the solution
corresponds to a flat $3$-brane in a $5D$ $AdS$ background,
found in \cite{rs}, and given by
$ds^2 = \exp(-\kappa^2_5 \sigma |w|/6)(-dt^2 + d\vec x^2) + dw^2$.
Hence we could look for a two-parameter family of solutions
which interpolates between these limits, which would be de-Sitter
on the wall and Anti-de-Sitter in the bulk.
However, rather then restricting
only to those $4D$ space-time sections which are de-Sitter,
we will merely require that they are maximally symmetric,
thus representing any of the de Sitter, Anti-de-Sitter or
Minkowski spacetimes in four dimensions.
The metric ansatz which accomplishes this is
\be
ds^2_5 = a^2(w) (-dt^2 + e^{2H t} d\vec x^2) + b^2(w) dw^2
\label{metans}
\ee
where the brane scale factor $H$ will be completely determined
by the cosmological terms $\Lambda$ and $\sigma$ after we
normalize the coefficient of $dt^2$ to unity on the brane.
Indeed, if $H$ is imaginary, $H = i{\cal H}$,
we can analytically continue the solution (\ref{metans})
by a coordinate transformation
$t = -i x'$, $x = it'$, $y =y'$ and $z = z'$
to
\be
ds^2_5 = a^2(w)
\Bigl(dx'^2 + e^{2{\cal H} x'}(-dt'^2 + dy'^2 + dz'^2)
\Bigr) + b^2(w) dw^2
\label{metansads}
\ee
a metric with $4D$ Anti-de-Sitter slices as claimed.
One must not ignore these solutions, since it is well known
\cite{hawk} that their slices correspond to open
FRW universes, and hence are not ruled out by observations.
In fact, for all cases where the brane metric is
locally $AdS_4$, we will
use the FRW representation
\be
ds^2_5 = a^2(w)
\Bigl(-dt^2 + |H|^{-2} \cos^2(|H|t)(\frac{dr^2}{1+r^2}
+ r^2 d\Omega)\Bigr) + b^2(w) dw^2
\label{frwads}
\ee
since we will have in mind cosmological evolution,
where the solution (\ref{frwads}) could be a reasonable
approximation at times after inflation on the brane.

To solve the equations of motion, we can substitute the ansatz
(\ref{metans}) into (\ref{ees}), and work out the details.
Alternatively, we can work in the action,
dimensionally reducing the theory to only one dimension,
which is possible since the $4D$ slices in (\ref{metans})
are maximally symmetric. The dependence of
all geometric quantities on the coordinates of $4D$ slices
is completely prescribed by the symmetry. 
A straightforward calculation
(see the appendix) gives
the equations of motion, which
in the gauge $b=1$ are
\ba
\label{eomsnew}
&&a'^2 = H^2 + \frac{\kappa^2_5 \Lambda}{6} a^2 \nonumber \\
&&a'' -\frac{\kappa^2_5 \Lambda}{6} a
=  -\frac{\kappa^{2}_5 }{3} \sigma a(0) \delta(w)
\ea
Again, we underline that $H^2$ can be both positive and
negative, despite the slightly confusing notation.
We will solve the equations for $H^2 > 0$, since the
solutions for $H^2<0$ can be easily obtained by
analytic continuation.
Leaving the details for the appendix, when $H^2 > 0$
we can write the solution as
\be
a = \frac{6 H}{\kappa_5
\sqrt{\kappa^2_5 \sigma^2 - 6\Lambda}} \Bigl(
\cosh(\sqrt{\frac{\kappa^2_5\Lambda}{6}}w)
- \frac{\kappa_5 \sigma}{\sqrt{6\Lambda}}
\sinh(\sqrt{\frac{\kappa^2_5 \Lambda}{6}} |w|) \Bigr)
\ee
We see that at the location of the brane, $a(0) = \frac{6H}{\kappa_5
\sqrt{\kappa^2_5 \sigma^2 - 6\Lambda}}$. We therefore normalize the
coordinates along the brane such that their coefficient is unity,
which is accomplished by
$t \rightarrow \kappa_5 
\frac{\sqrt{\kappa_5^2\sigma^2 - 6\Lambda}}{6H} t$,
and
$\vec x \rightarrow \kappa_5
\frac{\sqrt{\kappa_5^2\sigma^2 - 6\Lambda}}{6H} \vec x$.
Thus the expansion rate along the brane is
$H = \kappa_5 \sqrt{\kappa_5^2\sigma^2 - 6\Lambda}/6$,
in comoving units. 
The normalized
warp factor is
\be
a = \cosh(\sqrt{\frac{\kappa^2_5\Lambda}{6}}w)
- \frac{\kappa_5 \sigma}{\sqrt{6\Lambda}}
\sinh(\sqrt{\frac{\kappa^2_5 \Lambda}{6}} |w|)
\label{scalefac}
\ee
Substituting the
coordinate transformation into the metric we find
\be
ds^2_5 = \Bigl(\cosh(\sqrt{\frac{\kappa^2_5 \Lambda}{6}} w)
-\frac{\kappa_5 \sigma}{\sqrt{6\Lambda}}
\sinh(\sqrt{\frac{\kappa^2_5 \Lambda}{6}} |w|) \Bigr)^2
\Bigl(-dt^2 + e^{\kappa_5 \sqrt{\kappa_5^2\sigma^2 - 6\Lambda} t/3}
d \vec x^2 \Bigr) + dw^2
\label{finalsol}
\ee
This, finally, is our new solution for the case of a single brane.
The warp factor is a combination of exponentials while the wall
is curved. We will discuss its properties in the next section.
Here we should point out
one generic feature of wall models with localized energy density.
Rather interestingly, the square of the rate of expansion on the brane,
given by the Friedmann equation in the usual $4D$ cosmology,
is now
\be
H^2  = \kappa^2_5 \frac{{\kappa_5^2\sigma^2 - 6\Lambda}}{36}
\label{Hubble}
\ee
This equation is a further extension of
the generalized Friedmann equations of \cite{bin} because
it incorporates bulk contributions in addition to the brane terms.
When the bulk cosmological constant is zero, the expansion rate
is given by the brane tension (and not its square). When the tension
is zero, the expansion rate is equal to the square root of the bulk
cosmological constant, i.e. just the usual Friedmann rule.
In the general case when both terms are nonzero, 
the brane expansion rate
is a hybrid of these two formulas. As we have announced
above, for sufficiently large $\Lambda > \kappa^2_5 \sigma^2/6$
the ``expansion rate" is imaginary, which means
that the brane geometry is $AdS_4$ rather than $dS_4$.
The imaginary
numbers are removed by a proper analytical continuation,
discussed above. We will return to more examples of single brane
solutions in the next section. 

Let us turn to solutions describing two branes.
We will only consider solutions where the branes
are perfectly parallel. Such solutions can be described
by the action we have discussed above, which includes
an additional contribution
$\sim - \int_{\partial \bar M} d^4x \sqrt{g}\bar \sigma$.
We will approach the problem by taking
the solution (\ref{finalsol}) describing one brane in $AdS_5$
as a background, and will look for ways to fit the other brane in it,
adjusting its tension as required.
The backgrounds are warped products of the $4D$ spacetime and the
fifth dimension, implying that such
geometries can be foliated by a congruence
of slices which are conformal to the $3$-branes.
This allows us to find solutions by a simple pasting of the
probe brane. Here we will briefly review the derivation,
referring the reader to the appendix.

In order to match the background solution with the
long range fields of the probe brane, and insure that the
resulting array is a solution of the Einstein's equations,
we will start with the generalized Friedmann equation (\ref{Hubble})
for the source brane together with the first of eq.
(\ref{eomsnew}). If we combine these two equations, we find
\be
a'^2 = \frac{\kappa_5^4\sigma^2}{36} - \frac{\kappa^2_5\Lambda}{6}
+ \frac{\kappa^2_5 \Lambda}{6} a^2
\label{constat}
\ee
The RHS of this equation is not
always positive definite, which will
enable us to construct  new solutions of
parallel branes, different from the cases considered in \cite{rs}.
This equation shows that both branes have the same intrinsic
curvature, given by (\ref{Hubble}), although their tensions
are different. Away from the branes the original
solution (\ref{scalefac})
solves (\ref{constat})
and the brane expansion rate is given by (\ref{Hubble}).
This is just the gravitational field of the source brane.
The second order differential equation in this case contains
an additional Dirac $\delta$-function,
\be
a'' -\frac{\kappa^2_5 \Lambda}{6} a
=  -\frac{\kappa^{2}_5 }{3} \sigma a(0) \delta(w)
-\frac{\kappa^{2}_5 }{3} \bar \sigma a(w_c) \delta(w-w_c)
\label{twobreqtex}
\ee
The effect of the branes
is to compactify the bulk on the circle,
$w \sim w + w_c$, and further orbifold the circle
by $Z_2$ operation $w \sim -w$ in order to stabilize the solution.
Substituting the solution (\ref{scalefac}) into
(\ref{twobreqtex}), we obtain the matching condition
\be
\Bigl(\sqrt{\frac{2\kappa^2_5 \Lambda}{3}} +
\frac{\kappa^3_5 \sigma \bar \sigma}{3\sqrt{6\Lambda}}
\Bigr)\sinh(\sqrt{\frac{\kappa^2_5 \Lambda}{6}} w_c)
= \frac{\kappa^{2}_5 }{3} \Bigl(\sigma + \bar \sigma \Bigr)
\cosh(\sqrt{\frac{\kappa^2_5 \Lambda}{6}} w_c)
\label{matchcon}
\ee
The existence of solutions of this equation depends
on the signs of the brane tensions $\sigma, \bar \sigma$ and
the ratios $\frac{\sqrt{6\Lambda}}{\kappa |\sigma|}$,
$\frac{\sqrt{6\Lambda}}{\kappa |\bar \sigma|}$,
as discussed in the appendix. 
One can group those solutions into two main categories.

The first category of solutions is qualitatively new,
since both branes have positive tension.
For such solutions, the bulk cosmological constant must either
be at least infinitesimally larger than the square of the
larger of two tensions, which without loss of generality
we can take to be the source brane, with tension $\sigma$:
$6\Lambda > \kappa_5^2 \sigma^2$,
or infinitesimally smaller than the square of the 
smaller of two tensions, taken here as the probe brane 
tension: $6\Lambda < \kappa^2_5 \bar
\sigma^2$. By (\ref{Hubble}), this
means that the branes can never be completely flat unless
they are infinitely far apart. Indeed, the
branes are separated by a distance $w_c$ given by
\be
\tanh(\sqrt{\frac{\kappa^2_5 \Lambda}{6}} w_c) =
\frac{\kappa_5 \sqrt{6\Lambda}(\sigma + \bar \sigma)}{6 \Lambda
+ \kappa^2_5 \sigma \bar \sigma}
\label{distconthreet}
\ee
and we see that if $6\Lambda = \kappa_5^2 \sigma^2$,
$w_c \rightarrow \infty$. The warp factor on the probe brane is
\be
a(w_c) = \sqrt{\Bigl|\frac{6\Lambda -\kappa_5^2 \sigma^2}{6\Lambda-
\kappa^2_5 \bar \sigma^2}\Bigr|}
\label{warpppb}
\ee
The equation for the warp factor is easily obtained from
substituting the jump of the warp factor at the other brane
$a'^2/a^2(w_c) = \kappa^4_5 \bar \sigma^2/36$ into
eq. (\ref{constat}). Clearly,
the warp factor on the brane with smaller tension
is smaller. Hence this brane should be
the candidate for our world, since the warping
will lower the scales there. The amount of warping
and the brane separation are
completely specified by the values of the
cosmological terms, which as we will see
below is a generic property of the parallel
brane configurations. Hence this obviously suggests
an intricate relationship between the cosmological constant
problem and the mass hierarchy. We will return to this
interesting relationship later. When $6\Lambda
> \kappa^2_5 \sigma^2$, since 
both branes feel a net negative curvature, they
may be viewed as FRW $AdS_4$ universes. The metric
of the complete solution can therefore be written as
\be
ds^2_5 = a^2(w)
\Bigl(-dt^2 + \frac{36}{6\kappa^2_5 \Lambda - \kappa^4_5 \sigma^2}
\cos^2(\frac{\kappa_5 \sqrt{6\Lambda - \kappa^2_5 \sigma^2}}{6} t)
(\frac{dr^2}{1+r^2}
+ r^2 d\Omega)\Bigr) + b^2(w) dw^2
\label{parbraads}
\ee
where $a(w)$ is given by (\ref{scalefac}) 
with additional identification
in the bulk $w \sim w+w_c$, $w\sim - w$. 
In the case $6\Lambda < \kappa^2_5 \bar \sigma^2$, 
the longitudinal metric is replaced
by the de-Sitter one.
The solution for $a(w)$ for
$w<0$ is therefore a mirror 
image of the solution for $w>0$ around
the vertical axis at $w=0$, 
as it should be for orbifold constructions.
Note that $a(w)$ is cusping upwards at both branes, because both
branes have positive tension. This is possible because the solution
(\ref{scalefac}) for $a$ bounces in the bulk between
the two branes, as can be seen from the eq. (\ref{constat}).
The RHS can vanish between the branes, and $a$ can reexpand again.
This of course is induced by the negative curvature on the branes.
We will discuss phenomenological aspects of these solutions later.

The second category is comprised of the solutions where
one of the branes has a negative tension, whereas the other
has a positive tension. The solutions however fall into
three very different subcategories. In the first
case, if $\sqrt{\Lambda} = \frac{\kappa_5 \sigma}{\sqrt{6}} =
- \frac{\kappa_5 \bar \sigma}{\sqrt{6}}$, the
distance between the branes $w_c$ is arbitrary, as well
as the amount of warping at the brane
with negative tension (we remind the reader that we have
normalized the warp factor to unity at the
source brane). This is the solution given in \cite{rs},
describing two flat branes.
In the other subfamily of solutions, if the absolute value of the
negative tension is even slightly larger than the positive tension,
which is possible if the bulk cosmological constant is
slightly smaller than the square of the positive tension
(which implies that both branes are inflating, as
can be seen from (\ref{Hubble})), the distance between
the branes is not arbitrary any more, but is completely
specified by the cosmological terms:
\be
\tanh(\sqrt{\frac{\kappa^2_5 \Lambda}{6}} w_c)
= \frac{\kappa_5 \sqrt{6\Lambda}
(|\bar \sigma| - \sigma)}{\kappa^2_5
\sigma |\bar \sigma| - 6\Lambda}
\label{distcondonet}
\ee
Obviously, $w_c$ could be arbitrary only in the
simultaneous limit $\sigma = |\bar \sigma| 
= \frac{\sqrt{6\Lambda}}{\kappa_5}$.
The warp factor at the other brane is
\be
a(w_c) = \sqrt{\frac{\kappa_5^2 \sigma^2 
- 6\Lambda}{\kappa^2_5 |\bar \sigma|^2
-6\Lambda}}
\ee
The warp factor at the brane 
with negative tension is smaller than
unity, which is qualitatively the same as in the
solution of \cite{rs}. Unlike there, however,
the amount of warping is completely specified once
the cosmological terms are given.
The explicit form of the solutions is locally the same as
(\ref{metans}), with the warp factor given by (\ref{scalefac})
and with identifications 
$w \sim w+w_c$, $w \sim -w$. The warp factor
cusps upwards on the positive 
tension brane, and downwards on the
negative tension brane. The third subfamily of solutions
has the positive tension of one brane larger
than the absolute value of the 
negative tension of the other brane.
The bulk cosmological constant
must be larger than the square of the positive tension.
Again, even if the differences between the cosmological terms
are infinitesimal, the distance between the branes is
completely specified once the cosmological terms are known:
\be
\tanh(\sqrt{\frac{\kappa^2_5 \Lambda}{6}} w_c)
= \frac{\kappa_5 \sqrt{6\Lambda}
(\sigma - |\bar \sigma|)}{6 \Lambda-
\kappa^2_5 \sigma |\bar \sigma|}
\label{distcontwot}
\ee
Both branes feel a net negative 
curvature, however, and hence the
metric of the solution is formally the same as
(\ref{parbraads}), with different values of the
cosmological terms. The warp factor $a$ again
cusps upwards on the positive tension brane and downwards
on the negative tension brane.
However, the warp factor at the negative tension
brane is profoundly
different from the previous two cases: it is much larger
there than on the brane with the positive tension. Its value is
also uniquely specified by the cosmological terms:
\be
a(w_c) = \sqrt{\frac{6\Lambda -\kappa_5^2 \sigma^2}{6\Lambda-
\kappa^2_5 |\bar \sigma|^2}}
\ee
This can be understood as follows: this warp factor is in fact
a negative of the actual solution of the Einstein's equations.
Since the physics on either brane does not know about the sign
of $a$, we could simply ignore it, and instead write just the
absolute value of $a$. Then the flow of the warp factor is
opposite to the previous example, and hence the magnitude of
$a$ is naturally greater at the brane with negative tension.

The latter two subfamilies of solutions
suggest that the construction of \cite{rs} satisfying
the condition $\sigma = |\bar \sigma| 
= \frac{\sqrt{6\Lambda}}{\kappa_5}$
may be unstable to small perturbations
of the brane tensions.
Even a small mismatch between the brane tensions
and the bulk cosmological 
constant could offset the necessary amount
of warping, $a(w_c) \sim 10^{-15}$, regardless of where
the perturbation takes the solution.
However, once solutions belong to either of the two subfamilies
with mismatch already present, their stability under small
perturbations is restored. We will consider some explicit examples
later.

There also exist solutions of two parallel branes when
$\Lambda < 0$ (positive cosmological constant). This can be
seen from (\ref{matchcon}) and the fact that the hyperbolic functions
should be replaced by trigonometric ones. 
These solutions can be useful as toy models for inflationary model
building. Here we will list the solutions with their 
principal properties, as we did for the case when $\Lambda > 0$.
We leave the details for the
appendix, and a brief discussion of phenomenology of such
solutions for the Sec. IV. 

All the solutions with $\Lambda<0$ represent two parallel inflating 
branes separated by some bulk distance in the orbifold construction.
Using (\ref{Hubble}), we see that 
the expansion rate is given by $H^2 = \kappa^2 
\frac{\kappa^2 \sigma^2 + 6 |\Lambda|}{36}$, and 
there is a net positive cosmological 
constant on each brane.
The metric of the complete solution is
\be
ds^2_5 = \Bigl(\cos(\sqrt{\frac{\kappa^2_5 |\Lambda|}{6}} w)
-\frac{\kappa_5 \sigma}{\sqrt{6|\Lambda|}}
\sin(\sqrt{\frac{\kappa^2_5 |\Lambda|}{6}} |w|) \Bigr)^2
\Bigl(-dt^2 + e^{\kappa_5 \sqrt{\kappa_5^2\sigma^2 + 6|\Lambda|} t/3}
d \vec x^2 \Bigr) + dw^2
\label{inflparw}
\ee
with the orbifold conditions $w \sim w + w_c$ and $w \sim -w$.
The solutions are now divided into three categories. There are
two rather special cases, and the first is defined
by $\bar \sigma = -\sigma$. It corresponds to two
branes with tensions of the opposite sign. Both of these 
branes are inflating, as
should be expected from (\ref{Hubble}) and $\Lambda < 0$. The 
interbrane distance is still fixed, and is given by
\be
w_c = \sqrt{\frac{6}{\kappa^2_5 |\Lambda|}} \pi
\ee
Having normalized the warp factor to unity on the positive
tension brane (source brane), we find that at the negative
tension brane it is just $a(w_c) = -1$. Hence in this
particular case, the mass scales on both branes coincide.
The second special case has both tensions positive.
However, the
bulk cosmological constant  
is given by the product of the 
tensions, $6 |\Lambda| = \kappa^2_5 \sigma \bar \sigma$.
In this case, 
the interbrane distance is
\be
w_c = \sqrt{\frac{3}{2\kappa^2_5 |\Lambda|}} \pi
\ee
while the warp factor is
\be
a(w_c) = -\frac{\kappa_5 \sigma}{\sqrt{6|\Lambda|}}
\ee
Despite the fact that both branes have positive tension,
there is still a horizon between them. The reason is that the
brane with larger tension is more repulsive than the one
with smaller tension, and that the bulk cosmological 
constant is positive. The last category of solutions is 
the most generic, and includes all those configurations 
for which $\sigma + \bar \sigma \ne 0$ and 
$6 |\Lambda| \ne \kappa^2_5 \sigma \bar \sigma$. Therefore,
the solutions satisfy
\be
\tan(\sqrt{\frac{\kappa^2_5 |\Lambda|}{6}} w_c)
= \frac{\kappa_5\sqrt{6|\Lambda|}(\sigma + \bar \sigma )}{
\kappa^2_5 \sigma \bar \sigma -6 |\Lambda|}
\label{disttrig}
\ee
wherefore we can determine the interbrane distance. 
It is given by
\be
w_c = \sqrt{\frac{6}{\kappa^2_5 |\Lambda|}} \Bigl(
\tan^{-1}\Bigl(\frac{\kappa_5\sqrt{6|\Lambda|}(\sigma + \bar \sigma )}{
\kappa^2_5 \sigma \bar \sigma -6 |\Lambda|}\Bigr) + 2n \pi\Bigr)
\label{tandist}
\ee
where the contribution $2n\pi$ takes into account the periodicity
of the tangent function. 
The warp factor is then given by 
\be
a(w_c) = \zeta \sqrt{\frac{6|\Lambda| + \kappa^2_5 \sigma^2}
{6|\Lambda| + \kappa^2_5 \bar \sigma^2}}
\label{tanwarp}
\ee
where $\zeta = - sgn\{\cos(\sqrt{\frac{\kappa^2_5|\Lambda|}{6}} w_c)\}$.
Thus, there may or may not be a horizon between the branes, depending 
on the interplay of the three terms $\sigma$, $\bar \sigma$ and $|\Lambda|$.
The distance between the branes is however controlled
by the scale of the bulk cosmological constant,
$L \sim \sqrt{\frac{6}{\kappa^2_5 |\Lambda|}}$, 
since the inverse tangent function is always bounded by
unity. Hence in a de Sitter background
the scale of the interbrane distance is only weakly affected by the
bulk geometry, unlike in the $AdS$ background.
We will discuss these solutions further in Sec. IV.

\section{Isolated Domain Walls}

We now examine the solutions (\ref{finalsol})
in more detail. Let us begin with $\Lambda > 0$
(which corresponds to negative
bulk cosmological constant) 
and $\sigma > 0$ (positive brane tension).
In the limit $\Lambda \rightarrow 0$
we recover precisely the inflating $3$-brane solution of
\cite{kl}, which generalizes the inflating
domain walls \cite{vilipsi} in $4D$ space-time.
In the limit $6\Lambda = \kappa^2_5 \sigma^2$, we get the
Minkowski $3$-branes in $AdS_5$ background, found by
Randall and Sundrum \cite{rs}.
For any value of $\Lambda$ between these
two extremes, $0 \le \Lambda \le \frac{\kappa^2_5 \sigma^2}{6}$,
we find inflating $3$-branes in $AdS_5$ backgrounds.

Another notable property of the solutions with
$0 \le \Lambda \le \frac{\kappa^2_5 \sigma^2}{6}$ is the
existence and location of the bulk event horizon. As can be seen
from (\ref{finalsol}), the $3$-brane is surrounded with a
Rindler-type event horizon in the bulk, which is
located at the hypersurface $w ={\rm const}$ where the
metric coefficient $g_{00}$ vanishes. A simple calculation
shows that this is at
\be
w_{H} = \sqrt{\frac{3}{2 \kappa^2_5 \Lambda}}
\ln\Bigl(\frac{\kappa_5 \sigma+ \sqrt{6\Lambda}}
{\kappa_5 \sigma - \sqrt{6\Lambda}}\Bigr)
\label{rindhor}
\ee
In the limit $\Lambda \rightarrow 0$, this reproduces
$w_H = \frac{3M^3_5}{4\pi \sigma}$, found in \cite{kl}
for an inflating $3$-brane in a flat bulk. On the other
hand, when $\Lambda \rightarrow \kappa^2_5 \sigma^2/6$,
$w_H$ diverges. The presence
of the Rindler horizon in the bulk could be helpful
because of the cosmic no-hair theorem \cite{nohair},
which has two aspects in this context.
First, it pushes any bulk 
perturbations which occur near the brane
away from it. Second, it prevents perturbations at
distances larger than the Rindler horizon from ever
affecting the brane. From the vantage point of the brane
those perturbations behave as if they have fallen inside of
a black hole. As a result, any perturbations at distances
shorter than the Rindler horizon are pushed outside of it,
while the perturbations at longer distances are harmless
to begin with. On the other hand, as
the brane expansion rate
$H  = \kappa_5 \frac{\sqrt{{\kappa_5^2\sigma^2 - 6\Lambda}}}{6}$
decreases (which can be accomplished either by increasing
the bulk cosmological
constant $\Lambda$ or by decreasing the
brane tension $\sigma$),
the Rindler horizon moves farther and farther from
the brane. Hence if the bulk was initially homogeneous
over distances $\sim \frac{3M^3_5}{4\pi \sigma}$, it
may become homogeneous
over ever larger distances as $H$ approaches zero.
To see when this may happen, 
we should note that the time it takes
for perturbations near the brane
to get near the event horizon is roughly given by
$\tau \sim w_H$. Hence to fully benefit from the presence
of the Rindler horizon in the bulk during inflation on the
wall, the expansion rate should decrease slowly,
at timescales at least of order of $w_H$ or larger.
Otherwise the horizon would move
away from the brane faster than
the perturbations, trapping them forever near the brane.
Therefore we find a fundamental criterion for the slow roll
condition for inflation on the brane, which should
be satisfied to simultaneously
resolve both the brane and the bulk homogeneity problem
without fine tuning: the expansion rate on the brane
should change at timescales at least as large as $w_H$.
This is effectively similar to
having asymmetric inflation of
\cite{ahdkmr}.

As long as the
Poincare symmetry on the $3$-brane is unbroken,
the range of parameters for which
$\Lambda > \kappa^2_5 \sigma^2/6$ cannot
be reached by the rolling of the cosmological terms. This
is because for $\Lambda > \kappa^2_5 \sigma^2/6$,
the expansion
rate on the brane is imaginary, and
so the intrinsic geometry of the brane is
$AdS_4$, as can be seen from 
(\ref{Hubble}). We have seen that our
new solution with two branes 
(\ref{parbraads}) is precisely of this form.
Indeed, this should be expected
since if the brane tension $\sigma$ vanishes,
the brane must bend to conform to the bulk $AdS_5$ geometry,
becoming just an $AdS_4$ slice of it.
The brane tension $\sigma$ 
resists this, and the favored ground state
of the system depends on the 
rigidity of the brane relative to the
extrinsic curvature imposed by the bulk.
We will not study the mechanisms which can break the
Poincare symmetry on the brane, but will merely
assume that solutions with a definitive sign of $H^2$ are
at least locally minima of the Hamiltonian, and
are perturbatively separated from the solutions with the opposite
sign of $H^2$. However, it would be interesting to investigate
this in more detail. We note that the hyperbolic functions in
$g_{00}$ of (\ref{finalsol}) remain unchanged.
The location of the Rindler horizon (\ref{rindhor}) gets pushed
beyond infinity, in a manner of speaking, because $g_{00}$ never
vanishes for any value of $w$.

What happens if $\Lambda < 0$? As we have already discussed
in the previous section for the case of two parallel branes, 
this corresponds to the positive bulk
cosmological constant, in our 
conventions, leading to an 
even more vigorous inflation on $3$-branes than when $\Lambda > 0$.
Indeed, this remains true for a single brane case too.
The solution 
is formally the
same as (\ref{inflparw}), except that the only identification
is $w \sim -w$, and otherwise $w$ is unbounded. 
The expansion rate along the brane is now increased by the bulk
contributions, as expected. The metric however has become
periodic in the bulk, including the periodic distribution
of the Rindler horizons which appear at distances
$w_H + n \pi \sqrt{\frac{6}{\kappa^2_5 |\Lambda|}}$,
where the fundamental distance $w_H$ is given by
\be
w_H = \sqrt{\frac{6}{\kappa^2_5 |\Lambda|}} \tan^{-1}\Bigl(
\frac{\sqrt6|\Lambda|}{\kappa_5 \sigma}\Bigr)
\label{trigrindh}
\ee
A curious property of this formula is that the location
of the Rindler horizon adjacent 
to the brane oscillates as a function
of $\frac{\sqrt{|\Lambda|}}{\sigma}$, 
and can be arbitrarily close and
far from the brane.

Now we consider the solutions for the brane with a negative
tension, $\sigma = -|\sigma|$. When $\Lambda > 0$, the solutions
are still given by (\ref{finalsol}). Rather peculiarly,
the intrinsic geometry on the brane is still
inflating, even though the tension is negative, as can
be seen clearly from the generalized Friedmann equation
(\ref{Hubble}). However, the difference appears in the
absence of the Rindler horizon in the bulk. The coefficient
$g_{00}$ does not vanish at any hypersurface $w = {\rm const}$.
In contrast to the case $\sigma > 0$,
$\Lambda > \kappa^2_5 \sigma^2/6$, the brane is now attractive
for the probes in the bulk, rather than repulsive. This means
that any object which may be present in the bulk will fall onto
the brane in a finite amount of time. Such a rain of
particles from extra dimensions
further underscores the instability of the solution
and may impose strong phenomenological constraints on
model building.
It would be of interest to investigate it in more detail.

The negative tension branes 
appear generically even more unstable
if we take $\Lambda < 0$. The solution is given by
(\ref{inflparw}), but without the periodic
identification $w \sim w+w_c$. 
The brane is
still attractive for 
the particles in the bulk. However, now it is
surrounded by a {\it repulsive}
Rindler horizon, which limits the causal influence
of the brane on the bulk physics. Hence the phenomena on the brane
could exert influence in the bulk up to horizon
distances, but not beyond. This means that outside of the
horizon, there may be, for example, a highly inhomogeneous distribution
of particles which will all however fall on the brane in some
finite time, and possibly destroy any homogeneity produced by
inflation. Clearly, such a situation would be devastating. However,
as we know, in this case the distance of the horizon oscillates,
and can be arbitrarily large. One may therefore imagine
some physics on the brane reaching out into
the bulk to homogenize it during such phases.
Determining the possibility or impossibility of such
phenomena would obviously require scrutiny beyond the scope of
the present investigation, and we will not address it further.

\section{Parallel Walls}

We now return to the solutions which describe
two parallel branes. Our main interest is to explore the
new family where both branes have positive tension (\ref{parbraads}),
but each have FRW $AdS_4$ geometry.
The branes must be negatively curved because
the effective cosmological constant felt on each
brane is not completely cancelled. This is acceptable in principle
if the curvature is small, even if negative.
The $AdS_4$ is crucial for the existence
of this solution, since it enables the warp factor
(\ref{scalefac}) to bounce between the branes, which can be seen
from eq. (\ref{constat}). However, we 
will also reflect on the solutions
where one wall has positive tension and another negative. In
particular, the example where the branes have a small
net positive curvature is quite interesting, since both branes are
simultaneously undergoing inflation.

The bulk geometry in the direction
orthogonal to the brane is more complex than in the example of
the two brane solution of \cite{rs}. The deviation of the
warp factor  given by (\ref{scalefac}) takes into account
the bending of the branes in the bulk. Still, we can follow
closely the proposal of \cite{rs}, which posits that a physical
particle living on either brane couples to only the
$4D$ conformal metric on each brane, while the warp factor
only renormalizes the physical scales of the theory.
However, an additional
constraint must be considered.
Since in general there is an excess cosmological
constant on each brane, we must ensure that
the numbers needed to produce the $M_{Pl} - m_{EW}$ hierarchy
do not give an excessively large leftover cosmological constant
on the brane. As we have suggested before, this means that
there is a very intricate relationship between the cosmological
constant problem and the hierarchy problem.
One must therefore use these extended criteria to check the
phenomenological feasibility of the constructions
based on parallel branes.
In general, we will see that this is not satisfied in
the simple examples,
since the numbers needed to produce mass hierarchy fall
short of cancelling the cosmological constant to
experimentally permitted values.
Nevertheless, for a certain range
of admissible parameters which do lead to the
correct mass hierarchy, there is a very significant
reduction of the effective cosmological constant on
the branes. In a sense the picture which emerges
is a hybrid of the ideas of \cite{savas} and \cite{rs}.
To reduce the net cosmological constant on the brane,
the unification scale should be lower than the Planck scale
and the bulk should be large. 
So a part of the hierarchy is generated by raising the $4D$
Planck scale as in \cite{savas}.
The warping then produces
additional contribution to the mass hierarchy, 
lowering the fundamental scale down to $m_{EW}$.
Therefore although this mechanism does not  yet
solve the cosmological problem, it is still quite interesting.
It gives a possibility to have a very small but nonzero
cosmological constant on the brane.
There may exist other solutions describing two branes,
with some additional scalars in the bulk (such as the
moduli fields in string theory) which could produce a
more complicated running of the warp in the bulk, yielding a more
efficient cancellation of the cosmological constant on our
wall. Such scenarios may be useful for understanding
why the cosmological constant may be very small but nonvanishing.

Since the scale of the particle
dynamics varies non-monotonically in the bulk, the hierarchy
would have to come from the ratio of the warp
factors at different branes. Having normalized one of them to unity,
we can simply use the value of the other as given in eq. (\ref{warpppb}).
Then we recall the relationship
between the fundamental scale, given by $M_*$ and the Planck scale
$M_{Pl}$. Since $\kappa^2_5 = \frac{8\pi}{M^3_{*}}$ and
$\kappa^2_P = \frac{8\pi}{M^2_{Pl}}$, using the action (\ref{actdef})
one finds that \cite{rs}
\be
\frac{1}{2\kappa_P^2} = \frac{1}{2\kappa^2_5}
\int^1_0 d\vartheta w_c a^2(w_c \vartheta)
\label{planckmass}
\ee
Our wall should be the one with smaller
warp factor, therefore it is parameterized by $\bar \sigma$.
The net cosmological constant on our wall
must be very small. The net cosmological 
constant on the walls is a mismatch
between the bulk cosmological constant and $\sigma^2$, as
can be seen from (\ref{Hubble}):
$|H|^2 = \kappa^2_5 \frac{6\Lambda - \kappa^2_5 \sigma^2}{36}
= M^2_{Pl} \epsilon$ where $\epsilon = \frac{\lambda}{M^4_{Pl}}$
is a small dimensionless parameter.

Using (\ref{distconthreet}), we can 
get the distance between the branes in terms
of the leftover cosmological constant
$\epsilon$. To the lowest order, we 
can use the formula (\ref{distconthreet})
and $\sqrt{\frac{\kappa^2_5 \Lambda}{6}} w_c
\sim \frac{\kappa^2_5 \sigma}{6} w_c$. Numerically, this quantity
must be large in order to match the small $\epsilon$. After a little
algebra, we find
\be
e^{-\frac{\kappa^2_5 \sigma}{6} w_c} =
\Bigl(\frac{9 M^2_{Pl} \epsilon}{\kappa^4_5 \sigma^2}\Bigr)^{1/2}
\label{distsmall}
\ee
This formula gives the distance between the branes.
When we substitute the formula for the warp factor
(\ref{scalefac}) into (\ref{planckmass}), evaluate the
integral and use $\epsilon << 1$, we find that
\be
\kappa^2_P = \frac{\kappa^4_5 \sigma}{3}
\ee
From this equation, we can express 
$\sigma$ as a function of $M_*$ and
$M_{Pl}$, getting $\sigma = \frac{3M^6_*}{8\pi M_{Pl}^2}$.
Next, it is easy to check that the hierarchy constraint is
\be
a(w_c) = \frac{m_{EW}}{M_*}
\ee
Evaluating $a(w_c)$ we find $a(w_c) \sim
2\sqrt{\epsilon}(\frac{M_{Pl}}{M_*})^3$. Now we
can finally consider the numerical
relationships. Our equations translate to
the following formulas for $\epsilon$, $\bar \sigma$ and $\sigma$ in
terms of the mass scales $M_*$, $M_{Pl}$, $m_{EW}$:
\ba
\epsilon&=& {\frac{m^2_{EW} M^4_*}{4M_{Pl}^6}}
\nonumber \\
\bar \sigma &=& \frac{3m_{EW} M_*^5}{8\pi M_{Pl}^2}
\nonumber \\
\sigma&=& \frac{3M^6_*}{8\pi M^2_{Pl}}
\label{ccs}
\ea
The distance between the branes is
\be
w_c = \Bigl(\frac{M_{Pl}}{M_*}\Bigr)^2 M^{-1}_*
\ln\Bigl(\frac{4 M^2_*}{m^2_{EW}}\Bigr)
\label{distcc}
\ee
These equations show that in addition to the mass hierarchy
there also arises an interesting
hierarchy of the cosmological constants: the
residual cosmological constant $\lambda$ is smaller from the
tension of our wall $\bar \sigma$, which in turn is smaller
from the tension of the other wall $\sigma$. The cosmological
terms increase by factors of $\frac{m_{EW}}{M_*}$.
All the while, the distance between the branes may be much
larger than the unification length $M^{-1}_*$.
In fact, even this fact is a confirmation of the
induced hierarchy of cosmological constants: the
separation between branes is a bulk phenomenon,
controlled by the bulk scale $\sigma^{1/4}$, while the
phenomena on the brane are controlled by $\lambda^{1/4}$.
Hence the situation indeed represents an interesting
combination of elements of 
\cite{savas} and \cite{rs}. However, one
can verify that there is a lower bound on $\lambda$,
$\lambda \ge 10^{-32}(TeV)^4$, which is saturated when $M_* \sim TeV$.
This is too large by about 30 orders of magnitude.
Also, in this case $\sigma \sim \bar \sigma \sim \lambda$,
and the smallness of $\lambda$ becomes just the usual
fine tuning. The distance between the branes would
then be $w_c \sim 10^{15} cm$.
Nevertheless, generically $\lambda$ may still be considerably
smaller than the natural scale 
set by the brane tension on the hidden
wall. This reduction of the cosmological constant
is purely geometrical. Since the brane dwells
in the curved bulk, when it is located in the region of smaller
bulk curvature, it will bend less. The amount of bending
measures the effective cosmological constant on the wall, and
is determined by the location of the brane, which is controlled
by the interactions of the brane with the bulk and the hidden brane.
Hence the bulk geometry needed to produce the hierarchy of
particle physics scales could be useful to reduce the
severity of the fine tuning of cosmological constant.

The warp factor for the solutions with one positive
tension brane and one negative tension brane is very similar
to the case when both branes have 
positive tension. Thus it is clear
that as long as the warp factor is very small on the brane with
negative tension and
$\kappa^2_5 \sigma^2 = 6\Lambda 
- \frac{36 M^2_{Pl}}{\kappa^2_5} \epsilon
\sim 6 \Lambda$, the equations (\ref{ccs}) and (\ref{distcc}) would
remain correct, now corresponding to a {\it positive}
cosmological constant $\lambda = \epsilon M^4_{Pl}$ on our wall.
However, if we substitute for the scale $M_* \sim M_{Pl}$,
as has been suggested in \cite{rs}, we would get
$\lambda \sim  10^{-32} M^4_{Pl}$, i.e. a very large net
cosmological constant.

In the case of solutions with one positive
tension brane and one negative tension brane,
there exists another possibility for getting a tiny
$\lambda$. Suppose that the bulk cosmological constant
is very small, $6\Lambda << \kappa^2_5 \sigma^2$.
Then if $\lambda=H^2 M^2_{Pl}$ is taken as the
usual net cosmological constant on the wall, from (\ref{Hubble})
if $\sigma/M^4_* < 1$, also $\sigma^2/M^8_* << 1$.
Because this combination contributes to $\lambda$,
it may be possible to make $\lambda$ very small.
However, as can be seen from (\ref{planckmass}),
in this case $\Lambda = \frac{3M^9_*}{4\pi M^4_{Pl}}$.
From $H^2 = \epsilon M^2_{Pl}$ and 
$6\Lambda << \kappa^2_5\sigma^2$
we get $\sigma = \frac{3}{4\pi} 
\sqrt{\epsilon} M^3_* M_{Pl}$. Therefore
the condition 
$6\Lambda << \kappa^2_5\sigma^2$ translates into
$\frac{M_*}{M_{Pl}} << \epsilon^{1/6}$. Getting $\epsilon$ to
be small enough requires $M_*$ too small, 
$M_* \le 10^{-20} M_{Pl} \sim 100 MeV$.
Strictly speaking, these arguments show that the special
solution with two flat branes and $6\Lambda = \kappa^2_5 \sigma^2
= \kappa^2_5 \bar \sigma^2$ \cite{rs} is unstable in the sense that
even a small mismatch between the bulk and brane terms will
leave too large a cosmological constant on our wall.

The solutions with $\Lambda < 0$ are rather different 
from those with $\Lambda > 0$. If we consider the integral 
(\ref{planckmass}), for a large range of parameters it is well
approximated by 
\be
w_c\int^1_0 d\vartheta a^2(w_c \vartheta) 
\sim (1+ \frac{\kappa^2_5 \sigma^2}{6|\Lambda|}) w_c
\ee
This is because the warp factor is now oscillatory. 
Hence, the reduced Planck scale is (up to factors
of order unity)
\be
\kappa^2_P \approx \frac{6 \kappa^2_5|\Lambda|}
{(6|\Lambda| + \kappa^2_5 \sigma^2) w_c}
\label{dsplanck}
\ee
Using $H^2 = \frac{\lambda}{M^2_{Pl}}$, we can determine the
net cosmological constant on the walls in this limit. It is
\be
\lambda = \frac{4\pi}{3} \frac{|\Lambda|}{M^6_* w_c} M^4_{Pl}
\ee
Now, to put our solutions in context, we recall that if the
theory is really unified at the scale $M_*$, then the bulk
cosmological constant cannot exceed it by much. Hence,
$\Lambda \le M^5_*$. Then 
requiring that $\lambda$ is within the observationally allowed range
would require $w_c \ge 10^{120} \frac{|\Lambda|}{M^5_*} M^{-1}_*$.
Because of the periodicity of the distance formula (\ref{tandist}),
this equation can be satisfied in principle. But from 
(\ref{dsplanck}), taking the natural scale of the tension
not to exceed $M_*$ as well, we get $M_{Pl}^2 \sim w_c M^3_*$.
Since $M_* \ge 1 TeV$, the distance between the branes will
reduce the net cosmological constant by at most $32$ orders
of magnitude, when $M_*$ is exactly a $TeV$. Hence again the
net cosmological constant on the wall is bounded from below
by $\lambda \ge 10^{-32} (TeV)^4$, unless the bulk cosmological
constant is fine-tuned to an extremely low value. However,
the formula for the warp factor (\ref{tanwarp}) shows that 
in such a case further fine tuning of the wall tensions  
would be required to produce the mass hierarchy. 

However, we again stress that
it is important to note that although the
parallel brane solutions discussed here 
only partially succeed in simultaneously resolving the
mass hierarchy problem and reducing the cosmological
constant fine tuning problem, the decrease in the
net cosmological term can be nonnegligible.
This suggests interesting possibilities. It may happen that
there exist parallel brane solutions similar to (\ref{parbraads}) but
with a different behavior of the bulk energy density, and different
brane tensions, which could give more efficient screening of the
cosmological constant on our wall. 
For example, such solutions could be realized
in presence of additional bulk scalars, which can further modify the
``running" of the cosmological terms in the bulk, or it
may happen that the cancellation can be augmented
in the presence of more large internal dimensions.
We must note here however that the cosmological constant problem
in a universe comprising of a single brane need not be
as closely related to the mass hierarchy, since if a brane is
alone in the universe, the cosmological constant on it can be
tuned to an arbitrary value. Thus the cosmological constant problem
acquires its more conventional guise in the lone brane models.

\section{Conclusions}

It is widely believed that the unified theories of interactions which
contain quantum gravity should have more than four space-time dimensions.
Contrasting the conventional approach to additional dimensions which
are curled up at extremely short distances, the recent
proposal for explaining the hierarchy problem envisions a more
active role of the extra dimensions. The suggestion that there may be
large sub-millimeter dimensions \cite{savas} has opened up a new alley
of model-building, with many interesting applications in particle
physics. The understanding of cosmology in such models is still
at an early stage. There are several proposals for the description
of the early universe which may give a picture that can conform
with the standard Big Bang cosmology at low temperatures 
\cite{savas,ahdkmr}. However,
it is also possible that the extra dimensions may play other
roles, complementing the proposal of \cite{savas} as
discussed in \cite{rs}.

Here we have taken the view that
the extra dimensions could play a role which combines the
proposals of \cite{savas} and \cite{rs}. This view was illustrated
by deriving new large families of solutions of Einstein's equations
that describe single $3$-branes and pairs of $3$-branes,
which are located in a curved $5D$ world, and may themselves
also be curved. Since the intrinsic curvature of the walls
amounts to a net cosmological constant as viewed by an observer
pinned to the wall, it is important to check that the numerical values
of parameters in models with hierarchy do not give an excessively large
cosmological constant. For solutions which represent individual
$3$-branes, the net cosmological constant can always be tuned
down to an arbitrarily small value. Thus, while these solutions
are safe, they do not tell anything about the cosmological constant
problem.

For the solutions of two parallel $3$-branes,
the situation is more interesting. The hierarchy
in these solutions emerges because two walls are located
at different places in the curved bulk. Hence the
variation of the metric in the bulk from one wall
to another may renormalize the microphysical scales.
As a result, the walls will generically have to bend in
order to fit in the bulk geometry. The amount
of bending of each wall depends on its location,
and the cosmological constant on the wall is proportional
to its net curvature. Therefore, there will be a natural
hierarchy of cosmological constants in the solutions.
We have seen that the net cosmological constant on
our wall may become many orders of magnitude smaller
than its naturally expected magnitude. However in the
simple solutions we have considered, this geometric
softening of the cosmological constant is not sufficiently
strong to bring it down to $10^{-120} M^4_{Pl}$.
Nevertheless, the solutions show a tendency to
produce a net cosmological constant on our wall
which can be much lower than expected, while
still keeping the true wall tension large.
Perhaps there exist different solutions representing
curved walls where the screening of the cosmological
constant on our wall is more efficient.

We underline however that even if the mechanism
were successful, it would not have represented a
full solution to the cosmological constant problem.
Rather, it would still need some amount of tuning,
such as explaining why exactly our wall would be
located at a place in the bulk where the bulk curvature
would bend it by a right amount. But this question
may be related to the problem of moduli stabilization, and
hence beyond the scope of the formalism pursued here.
However, if one assumes that the moduli can be stabilized,
then the location of the brane in the bulk may help
explain why cosmological constant would be small
but nonzero.

The cosmology of models with dynamical extra dimensions
is still at a very early stage, and various investigations
have been done so far \cite{bendav,kl,dt,ahdkmr,ovst,low,bin,lyth,
edi,banks,othercosmo,dvali}. Here we have considered a simple
effective description of domain walls in $5D$, and have found
that among the curved domain walls there is a profusion of
those which are inflating in the 
longitudinal directions. But before
such solutions can be viewed as 
candidates for realistic models
of brane cosmology, it must be 
possible for inflation on the brane
to end, accomplishing many set requirements, such as
producing enough density perturbations and sufficient reheating.
Here we will briefly remark on some basic features of inflationary
model building in the braneworld framework. The examples
of inflating walls show that the wall inflation, driven by some
conventional inflationary model on the wall which controls
the evolution when the bulk physics 
is weakly coupled, could also help with the
horizon problem in the bulk \cite{kl}. A conventional
inflation produces a positive cosmological constant
along the wall, which means that the wall will have positive tension.
Because gravity in $5D$ will not be totally confined to the wall,
and during inflation driven by a positive tension,
the wall's gravitational field will repulse any objects in the
bulk. This is a well known property of domain walls in
flat $4D$ spacetime \cite{vilipsi}, and all solutions
with positive tension also share it. This could solve the bulk
horizon problem as follows. When the expansion of the brane is fast,
both the brane and bulk horizons are small, close to the fundamental
scale. There will be many such 
sufficiently small causally isolated domains
which are almost smooth both along the brane and around it.
They can therefore start to inflate. Hence if
inflation on the wall lasts long enough and proceeds slowly, the
gravitational field of the wall will push out inhomogeneities
in the bulk very far away from the wall.
In this fashion, the bulk around 
the wall can also become homogeneous
during wall inflation. 
Another interesting aspect is the interplay of positive bulk
and brane cosmological terms, which can rapidly 
inflate only the wall, while leaving the bulk size 
almost constant. The resulting inflationary expansion 
is then highly asymmetric, of precisely the type discussed
in the longwavelength limit in \cite{ahdkmr}.
It may be necessary that in order to realize this
possibility the bulk degrees of freedom
must be very light, or very weakly coupled \cite{bendav,lyth,kl}. 
However, such models
may emerge in theories which have large extra dimensions,
and thus this is certainly a possibility \cite{dt,edi}.

Writing down a concrete model is not as easy as it might seem,
however. If the large internal dimensions are stabilized with the 
bulk size
smaller than the horizon length on the wall,
the approximation based on applying the usual $4D$ scenarios
and ignoring the gravitational effects in the extra dimensions
is a good one, and scaled-down versions of models such as
\cite{andc}-\cite{Dterm} may apply. 
If the extra dimensions are dynamical,
it may happen that the $4D$ scenarios 
are still applicable \cite{ahdkmr}.
In other scenarios
the effects of gravity in the extra dimensions may be
significant, as we have seen here (see also \cite{bin}), and
may require a different approach. An example we have noted
here is the interplay of the wall and the bulk physics via
gravity, manifest in the role of the bulk event horizon.
This indicates that higher-dimensional
dynamics in a generic situation warrants 
further investigation.

{\it {\bf Note added in proof:} While this paper was being submitted,
we became aware of the work \cite{nih}, which contains a subset
of the solutions presented here.}

\vspace{.5cm}
{\bf Acknowledgements}

We would like to thank S. Dimopoulos, E. Halyo,
A. Linde and M. Luty for useful conversations.
This work has been supported in part by
NSF Grant PHY-9870115.

\appendix

\section{Appendix}

We are looking for the solutions of the form (\ref{metans}).
For the wall de-Sitter metric we chose the parametrization
$dS^2_{dS} = \frac{-d\tau^2 + d\vec x^2}{H^2 \tau^2}$.
These coordinates are related to familiar
de-Sitter coordinates  by $\tau = \kappa^{2/3}_5 \exp(-Ht)$,
$\vec x = \kappa^{2/3}_5 H \vec y$.
The brane curvature $H$ will be determined by
the cosmological terms $\Lambda$ and $\sigma$, modulo a
coordinate transformations. We will see that later.
The 5D metric of interest is
\be
\label{5dmetric}
ds^2 = a^2(w) \frac{-d\tau^2 + d \vec x^2}{H^2 \tau^2} + b^2(w) dw^2
\ee
To get the reduction, we can evaluate the Ricci scalar for
this metric.
A straightforward calculation gives
\be
\label{ricci}
R = 12\frac{H^2}{a^2} - 12 \frac{a'^2}{a^2 b^2} - \frac8{ab}
(\frac{a'}{b})'
\ee
By our ansatz,
$\sqrt{g}(bulk) = a^4 b/(H \tau)^4$ and
$\sqrt{g}(boundary) = a^4/(H \tau)^4$,
and so the volume of the de-Sitter space
is $V_4=\int \frac{d\tau d^3 \vec x}{H^4 \tau^4}$.
As is usual in supergravity models, we
add to the action the usual Gibbons-Hawking term, which
precisely cancels a total derivative term.
Therefore the reduced effective action
$I = 2 \kappa^2_5 S /V_4$ is
\be
\label{redeffact}
I = \int dw \Bigl\{12 H^2 a^2 b + 12 \frac{a^2 a'^2}{b}
+ 2\kappa^2_5 \Lambda b a^4 \Bigr\}
-2\kappa^{2}_5 \int dw a^4 \sigma \delta(w)
\ee
where we have introduced the Dirac $\delta$
functions to represent the boundary terms
by a fictitious bulk Lagrangian, in order
to facilitate the variational procedure.
To get the equations of motion, we now
vary this action with respect to
$b$ and $a$. After some simple algebra,
with the gauge $b=1$, the equations become
\be
\label{eomsnewn}
a'^2 = H^2 + \frac{\kappa^2_5 \Lambda}{6} a^2 ~~~~~~~~~~
a'' -\frac{\kappa^2_5 \Lambda}{6} a
=  -\frac{\kappa^{2}_5 }{3} \sigma a(0) \delta(w)
\ee
These equations are actually easy to solve.
We can readily verify that
the solution must be of the form
\be
a = \alpha \sinh(\sqrt{\frac{\kappa^2_5 \Lambda}{6}} |w|)
+ \beta \cosh(\sqrt{\frac{\kappa^2_5\Lambda}{6}}w)
\label{scalefaca}
\ee
to solve the second order differential equation.
It is useful to recall that
\ba
\sinh(\sqrt{\frac{\kappa^2_5 \Lambda}{6}} |w|)
&=& (2\theta(w)-1) \sinh(\sqrt{\frac{\kappa^2_5 \Lambda}{6}} w)
\nonumber \\
(\sinh(\sqrt{\frac{\kappa^2_5 \Lambda}{6} } |w|))' &=&
\sqrt{\frac{\kappa^2_5 \Lambda}{6}}
\cosh(\sqrt{\frac{\kappa^2_5 \Lambda}{6}}w)(2\theta(w) -1)
\nonumber \\
(\sinh(\sqrt{\frac{\kappa^2_5 \Lambda}{6}} |w|))''
&=& \sqrt{\frac{2\kappa^2_5 \Lambda}{3}}\delta(w)
+ {\frac{\kappa^2_5 \Lambda}{6} }
\sinh(\sqrt{\frac{\kappa^2_5 \Lambda}{6}}|w|)
\ea
where $\theta(w)$ is the step function.
Then using this, we find that $a''
- \frac{\kappa^2_5 \Lambda}{6} a =
\sqrt{\frac{2\kappa^2_5 \Lambda}{3}} \alpha \delta(w)$
and hence the
coefficients $\alpha$ and $\beta$ must satisfy
\be
\alpha = -\frac{\kappa_5 \sigma}{\sqrt{6\Lambda}} \beta
\label{coeffone}
\ee
This is the first relationship. The second relationship
comes from the constraint equation: since
$a'^2 - \frac{\kappa^2_5 \Lambda}{6} a^2
= H^2$, we get
\be
\alpha^2 - \beta^2 = \frac{6 H^2}{\kappa^2_5 \Lambda}
\label{coefftwo}
\ee
Using (\ref{coeffone}) and (\ref{coefftwo}),
we can evaluate $\alpha$ and $\beta$,
and find
\be
\beta = \frac{6 H}{\kappa_5
\sqrt{\kappa^2_5 \sigma^2 - 6\Lambda}} ~~~~~~~~~~
\alpha = -
\frac{\sqrt{6} \sigma H}{\sqrt{\Lambda}
\sqrt{\kappa^2_5 \sigma^2 - 6\Lambda}}
\ee
Therefore, the expression for the warp factor is
\be
a = \frac{6 H}{\kappa_5
\sqrt{\kappa^2_5 \sigma^2 - 6\Lambda}} \Bigl(
\cosh(\sqrt{\frac{\kappa^2_5\Lambda}{6}}w)
- \frac{\kappa_5 \sigma}{\sqrt{6\Lambda}}
\sinh(\sqrt{\frac{\kappa^2_5 \Lambda}{6}} |w|) \Bigr)
\ee
At the brane, $a(0) = \frac{6H}{\kappa_5
\sqrt{\kappa^2_5 \sigma^2 - 6\Lambda}}$, and we can normalize the
coordinates along the brane such that their coefficient is unity,
by the coordinate transformations
$t \rightarrow \kappa_5 
\frac{\sqrt{\kappa_5^2\sigma^2 - 6\Lambda}}{6H} t$,
and
$\vec x \rightarrow \kappa_5
\frac{\sqrt{\kappa_5^2\sigma^2 - 6\Lambda}}{6H} \vec x$,
and the expansion rate along the brane 
$H = \kappa_5 \sqrt{\kappa_5^2\sigma^2 - 6\Lambda}/6$
in comoving units. 
The normalized warp factor is
\be
a = \Bigl(
\cosh(\sqrt{\frac{\kappa^2_5\Lambda}{6}}w)
- \frac{\kappa_5 \sigma}{\sqrt{6\Lambda}}
\sinh(\sqrt{\frac{\kappa^2_5 \Lambda}{6}} |w|) \Bigr)
\ee
When $\Lambda < 0$ (positive cosmological constant), the hyperbolic
functions in (\ref{matchcona})
should be replaced by trigonometric ones. 
In this case, the warp factor is, after normalizing it to unity
on the brane with tension $\sigma$,
\be
a = \Bigl(
\cos(\sqrt{\frac{\kappa^2_5|\Lambda|}{6}}w)
- \frac{\kappa_5 \sigma}{\sqrt{6|\Lambda|}}
\sin(\sqrt{\frac{\kappa^2_5 |\Lambda|}{6}} |w|) \Bigr)
\label{warptriga}
\ee
and the expansion rate is 
$H^2 = \kappa^2_5 \frac{\kappa^2_5 \sigma^2 + 6|\Lambda|}{36}$. 
This completes our search for
the single brane solutions.

Let us now consider parallel branes.
We will look for the solution with two branes by taking
(\ref{finalsol}) which describes one brane in $AdS_5$
as a background, and will fit the other brane in it,
adjusting its tension as required.
We will only consider the cases
where the bulk geometry can be foliated by a congruence
of slices which are conformal to the $3$-branes.
Then away from the branes the original
solution (\ref{scalefaca})
would continue to solve the first of the eqs.
(\ref{eomsnewn}), which explicitly gives
\be
a'^2 = \frac{\kappa_5^4\sigma^2}{36} - \frac{\kappa^2_5 \Lambda}{6}
+ \frac{\kappa^2_5 \Lambda}{6} a^2
\label{consta}
\ee
for $a$ given by (\ref{scalefaca}) and with $\beta =1$,
$\alpha = -\kappa_5 \sigma/\sqrt{6\Lambda}$ (which corresponds
to the brane expansion rate
$H = \kappa_5 \sqrt{\kappa^2_5 \sigma^2 -6\Lambda}/6$).
The second equation in this case is
\be
a'' -\frac{\kappa^2_5 \Lambda}{6} a
=  -\frac{\kappa^{2}_5 }{3} \sigma a(0) \delta(w)
 -\frac{\kappa^{2}_5 }{3} \bar \sigma a(w_c) \delta(w-w_c)
\label{twobreq}
\ee
It is evident that (\ref{scalefaca}) will satisfy this
equation away from the branes, where $\delta$-functions vanish.
To take into account the effect of the branes,
we compactify the bulk on the circle,
$w \sim w + w_c$, and further orbifold the circle
by $Z_2$ operation $w \sim -w$. Then we can substitute
(\ref{scalefaca}) into (\ref{twobreq})
and after some simple algebra find
the required matching condition:
\be
\Bigl(\sqrt{\frac{2\kappa^2_5 \Lambda}{3}} +
\frac{\kappa^3_5 \sigma \bar \sigma}{3\sqrt{6\Lambda}}
\Bigr)\sinh(\sqrt{\frac{\kappa^2_5 \Lambda}{6}} w_c)
= \frac{\kappa^{2}_5 }{3} \Bigl(\sigma + \bar \sigma \Bigr)
\cosh(\sqrt{\frac{\kappa^2_5 \Lambda}{6}} w_c)
\label{matchcona}
\ee
The solutions of (\ref{matchcona}) are
controlled by the signs of $\sigma, \bar \sigma$ and
the ratios $\frac{\sqrt{6\Lambda}}{\kappa |\sigma|}$,
$\frac{\sqrt{6\Lambda}}{\kappa |\bar \sigma|}$.
Bearing in mind that $w_c > 0$ (we ignore branes
lying on top of each other since they are indistinguishable
from single branes in our approach), and focusing first on $\Lambda>0$
case, it is straightforward
to verify that solutions  exist in the following cases:

$\bullet$ if $\sigma, \bar \sigma > 0$ and
$6\Lambda > {\rm max}\{\kappa^2_5 \sigma^2, \kappa^2_5 \bar \sigma^2\}$
or $6 \Lambda < {\rm min} \{\kappa^2_5 \sigma^2, \kappa^2_5 \bar
\sigma^2\}$.
This solution is qualitatively different from examples
considered by \cite{rs}.
The distance between the branes is
\be
\tanh(\sqrt{\frac{\kappa^2_5 \Lambda}{6}} w_c)
= \frac{\kappa_5 \sqrt{6\Lambda}(\sigma + \bar \sigma)}{6 \Lambda +
\kappa^2_5 \sigma \bar \sigma}
\label{distconthree}
\ee
while the warp factor on the other brane is
\be
a(w_c) = \sqrt{\Bigl|\frac{6\Lambda -\kappa_5^2 \sigma^2}{6\Lambda-
\kappa^2_5 \bar \sigma^2}\Bigr|}
\ee
In this case, the warp factor on the other brane can be both larger
or smaller than unity, depending on the ratio of the tensions
$\sigma/\bar \sigma$. If they are equal, then the warp factor
on both branes is exactly unity, and the branes are at a distance
$\tanh(\sqrt{\frac{\kappa^2_5 \Lambda}{6}} w_c) =
\frac{2\sqrt{6\Lambda} \kappa_5 \sigma}{6\Lambda + \kappa_5 \sigma^2}$.

$\bullet$ either $\sqrt{\Lambda} = 
\frac{\kappa_5 \sigma}{\sqrt{6}}$ or
$\sqrt{\Lambda} = 
\frac{\kappa_5 \bar \sigma}{\sqrt{6}}$; for both
cases the constraint (\ref{matchcona}) reduces to
$(\sigma + \bar \sigma) 
\sinh(\sqrt{\frac{\kappa^2_5 \Lambda}{6}} w_c)
= (\sigma + \bar \sigma) 
\cosh(\sqrt{\frac{\kappa^2_5 \Lambda}{6}} w_c)$.
The only solutions of this equation are $\bar \sigma = - \sigma$,
when the distance between the branes $w_c$ can be arbitrary,
which is precisely the solution of \cite{rs},
or $w_c \rightarrow \infty$ if $\sigma + \bar \sigma \ne 0$, which
reduces to a single brane solution.

$\bullet$ if $\sigma > 0$ and $\bar \sigma < 0$ 
but $\sigma + \bar \sigma<0$,
there exist solutions when 
$6\Lambda < \kappa_5^2 \sigma^2$. The distance
between the two branes is determined by the equation
\be
\tanh(\sqrt{\frac{\kappa^2_5 \Lambda}{6}} w_c)
= \frac{\kappa_5 \sqrt{6\Lambda}(|\bar \sigma| - \sigma)}{\kappa^2_5
\sigma |\bar \sigma| - 6\Lambda}
\label{distcondone}
\ee
Note that the solution converges to that of \cite{rs} in the
simultaneous limit $\sigma = |\bar \sigma| 
= \frac{\sqrt{6\Lambda}}{\kappa_5}$.
The warp factor on the brane with negative tension
\be
a(w_c) = \sqrt{\frac{\kappa_5^2 \sigma^2 
- 6\Lambda}{\kappa^2_5 |\bar \sigma|^2
-6\Lambda}}
\ee
and is smaller than unity. In the limit
$\sigma = |\bar \sigma| = \frac{\sqrt{6\Lambda}}{\kappa_5}$
the ratio becomes undetermined. This is equivalent to the fact
that the distance between the 
branes is arbitrary in this specific limit.

$\bullet$ if $\sigma > 0$, $\bar \sigma <0$ 
but $\sigma + \bar \sigma > 0$,
the solutions exist if $6\Lambda > \kappa^2_5 \sigma^2$.
In this case, the distance between the branes is
\be
\tanh(\sqrt{\frac{\kappa^2_5 \Lambda}{6}} w_c)
= \frac{\kappa_5 \sqrt{6\Lambda}(\sigma - |\bar \sigma|)}{6 \Lambda-
\kappa^2_5 \sigma |\bar \sigma|}
\label{distcontwo}
\ee
while the warp factor at the location of the second brane is
\be
a(w_c) = \sqrt{\frac{6\Lambda -\kappa_5^2 \sigma^2}{6\Lambda-
\kappa^2_5 |\bar \sigma|^2}}
\ee
Note that in this case the warp factor 
at the negative tension brane
is greater than unity. Once more, in the limit
$\sigma = |\bar \sigma| = \frac{\sqrt{6\Lambda}}{\kappa_5}$
both the warp factor and the distance between the branes become
undetermined.

We now consider parallel branes when $\Lambda < 0$. 
All of these solutions
can be thought of
as two parallel inflating 
branes in the orbifold construction.
Again, by the orbifolding condition requres $w_c >0$. 
Then the matching 
condition (\ref{matchcona}) is replaced by
\be
\Bigl(\frac{\kappa^3_5 \sigma \bar \sigma}{3\sqrt{6|\Lambda|}}
-\sqrt{\frac{2\kappa^2_5 |\Lambda|}{3}}
\Bigr)\sin(\sqrt{\frac{\kappa^2_5 |\Lambda|}{6}} w_c)
= \frac{\kappa^{2}_5 }{3} \Bigl(\sigma + \bar \sigma \Bigr)
\cos(\sqrt{\frac{\kappa^2_5 |\Lambda|}{6}} w_c)
\label{matchcontriga}
\ee
The solutions of this equation can be divided into three categories:

$\bullet$ if $\bar \sigma = - \sigma$, we must have 
$\sin(\sqrt{\frac{\kappa^2_5 |\Lambda|}{6}} w_c) = 0$, 
leading to the interbrane distance 
\be
w_c = \sqrt{\frac{6}{\kappa^2_5 |\Lambda|}} \pi
\ee In 
this case, the warp factor on the brane with the negative
tension is $a=-1$. The sign change shows that there is 
a Rindler horizon between the branes. Its precise location is given by
\be
w_H = \sqrt{\frac{6}{\kappa^2_5 |\Lambda|}} \tan^{-1}
\Bigl(\frac{\kappa_5 \sigma}{\sqrt{6|\Lambda|}}\Bigr)
\ee
The brane with the negative tension attracts every object in the
bulk, whereas the brane with the positive tension repells them,
although both are exponentially inflating.

$\bullet$ if $\sigma > 0$, $\bar \sigma > 0$ and 
$6 |\Lambda| = \kappa^2_5 \sigma \bar \sigma$, the solutions
must satisfy $\cos(\sqrt{\frac{\kappa^2_5 |\Lambda|}{6}} w_c) = 0$,
therefore giving for the interbrane distance
$w_c = \sqrt{\frac{3}{2\kappa^2_5 |\Lambda|}} \pi$. The warp factor on the
other brane is 
\be
a(w_c) = -\frac{\kappa_5 \sigma}{\sqrt{6|\Lambda|}}
\ee
Hence there is still a horizon between the branes, although
both have positive tension. The reason for this is that the
brane with larger tension is more repulsive than the one
with smaller tension. 

$\bullet$ if $\sigma + \bar \sigma \ne 0$ and 
$6 |\Lambda| \ne \kappa^2_5 \sigma \bar \sigma$, the solutions must
satisfy
\be
\tan(\sqrt{\frac{\kappa^2_5 |\Lambda|}{6}} w_c)
= \frac{\kappa_5\sqrt{6|\Lambda|}(\sigma + \bar \sigma )}{
\kappa^2_5 \sigma \bar \sigma -6 |\Lambda|}
\label{disttriga}
\ee
This equation determines the separation between the 
branes. It is given by 
\be
w_c = \sqrt{\frac{6}{\kappa^2_5 |\Lambda|}} \Bigl(
\tan^{-1}\Bigl(\frac{\kappa_5\sqrt{6|\Lambda|}(\sigma + \bar \sigma )}{
\kappa^2_5 \sigma \bar \sigma -6 |\Lambda|}\Bigr) + 2n\pi \Bigr)
\ee
The warp factor is then given by 
\be
a(w_c) = \zeta \sqrt{\frac{6|\Lambda| + \kappa^2_5 \sigma^2}
{6|\Lambda| + \kappa^2_5 \bar \sigma^2}}
\ee
where $\zeta = - sgn\{\cos(\sqrt{\frac{\kappa^2_5|\Lambda|}{6}} w_c)\}$.
Thus, there may or may not be a horizon between the branes, depending 
on the interplay of the three terms $\sigma$, $\bar \sigma$ and $|\Lambda|$.


\begin{thebibliography}{99}

\bibitem{savas} N. Arkani-Hamed, S. Dimopoulos and G. Dvali,
{\it Phys. Lett.} {\bf B429}, 263 (1998); hep-ph/9807344;
I. Antoniadis, N. Arkani-Hamed, S. Dimopoulos and G. Dvali,
{\it Phys. Lett.} {\bf B436}, 257 (1998).

\bibitem{ant} I. Antoniadis, \PL {\bf B246}, 377 (1990);
I. Antoniadis and K. Benakli, \PL {\bf B326} 69 (1994);
I. Antoniadis, K. Benakli and M. Quiros, \PL {\bf B331} 313 (1994).

\bibitem{ahdmr} N. Arkani-Hamed, S. Dimopoulos and
J. March-Russell, hep-th/9809124.

\bibitem{dienes} K. Dienes, E. Dudas
and T. Gherghetta, {\it Phys. Lett.} {\bf B436}, 55 (1998);
hep-ph/9806292; hep-ph/9807522;
K. Dienes, E. Dudas, T. Gherghetta and
A. Riotto, hep-ph/9809406.

\bibitem{others} R. Sundrum, hep-ph/9805471; hep-ph/9807348;
G. Shiu and S.-H. Tye, {\it Phys. Rev.} {\bf D58}, 106007 (1998);
Z. Kakushadze and S.-H. Tye,   hep-th/9809147;
C. Bachas, hep-ph/9807415; K. Benakli, eprint
hep-ph/9809582; L. Randall and
R. Sundrum,  hep-th/9810155.

\bibitem{bendav} K. Benakli and S. Davidson, hep-ph/9810280.

\bibitem{hw} P. Ho\v rava and
E. Witten, {\it Nucl. Phys.} {\bf B460}, 506 (1996);
E. Witten, {\it Nucl. Phys.} {\bf 471}, 135 (1996).

\bibitem{string}
J.D. Lykken, {\it Phys. Rev.} {\bf D54}, 3693 (1996);
I. Antoniadis and M. Quiros,
{\it Phys. Lett.} {\bf B392}, 61 (1997);
C.P. Burgess, L.E. Ibanez and F. Quevedo,   hep-ph/9810535;
L.E. Ibanez, C. Munoz and S. Rigolin, hep-ph/9812397;
I. Antoniadis, G. D'Appollonio, E. Dudas and A. Sagnotti,
hep-th/9812118.

\bibitem{new}
G.F. Giudice, R. Rattazzi and J.D. Wells, hep-ph/9811291;
M. Maggiore and A. Riotto, hep-th/9811089;
S. Nussinov and R. Shrock, hep-ph/9811323;
E.A. Mirabelli, M. Perelstein and M.E. Peskin, hep-ph/9811337;
T. Han, J.D. Lykken and R. Zhang, hep-ph/9811350;
J.L. Hewett, hep-ph/9811356;
Z. Berezhiani and G. Dvali, hep-ph/9811378;
K.R. Dienes, E. Dudas and T. Gherghetta, hep-ph/9811428;
N. Arkani-Hamed, S. Dimopoulos, G. Dvali
and J. March-Russell, hep-ph/9811448;
Z. Kakushadze, hep-th/9811193; hep-th/9812163;
I. Antoniadis and C. Bachas, hep-th/9812093;
S. Cullen and M. Perelstein, hep-ph/9903422.

\bibitem{kl} N. Kaloper and A. Linde, hep-th/9811141.

\bibitem{dt}
G. Dvali and S.H.H. Tye, hep-ph/9812483.

\bibitem{ahdkmr}
N. Arkani-Hamed, S. Dimopoulos, N. Kaloper and J. March-Russell,
hep-ph/9903224; hep-ph/9903239.


\bibitem{rus} V. Rubakov and M. Shaposhnikov,
{\it Phys. Lett.} {\bf B125}, 136 (1983);
A.T. Barnaveli and O.V. Kancheli,
{\it Sov. J. Nucl. Phys.} {\bf 52}, 576 (1990).

\bibitem{ds} G. Dvali and M. Shifman,
{\it Nucl. Phys.} {\bf B504}, 127 (1997).

\bibitem{rs}
L. Randall and R. Sundrum, hep-ph/9905221.

\bibitem{fr}
P.G.O. Freund and M.A. Rubin, \PL {\bf B97} 233 (1980);
F. Englert, \PL {\bf B119} 339 (1982); 
M.J. Duff, B.E.W. Nilsson and C.N. Pope, \PRL {\bf 50} 2043 (1983).

\bibitem{iib}
J. Maldacena, {\it Adv. Theor. Math. Phys.} {\bf 2} 231 (1998);
P. Claus, R. Kallosh, J. Kumar, P. Townsend and A. Van Proeyen,
{\it JHEP} {\bf 9806} 004 (1998).

\bibitem{sugrwalls}
H. Lu, C.N. Pope and P.K. Townsend, \PL {\bf B391} 39 (1997);
P.M. Cowdall, H. Lu, C.N. Pope, K.S. Stelle and
P.K. Townsend, 
\NP {\bf B486} 49 (1997);
M.S. Bremer, M.J. Duff, H. Lu, C.N. Pope and K.S. Stelle,
\NP {\bf B543} 321 (1999);
M. Cvetic,  J.T. Liu, H. Lu and C.N. Pope,  hep-th/9905096.


\bibitem{cvetic}
M. Cvetic, S. Griffies and H. Soleng, \PR {\bf D48} 2613 (1993);
M. Cvetic and H. Soleng, {\it Phys. Rept.} {\bf 282} 159 (1997).

\bibitem{ovst}
A. Lukas, B.A. Ovrut, K.S. Stelle and D. Waldram,
\PR {\bf D59}, 086001 (1999).

\bibitem{low}
A. Lukas, B.A. Ovrut and D. Waldram, hep-th/9902071.

\bibitem{hawk}
S.W. Hawking and G.F.R. Ellis, {\it The large scale structure of
space-time}, Cambridge University Press, Cambridge 1973.

\bibitem{bin}
P. Binetruy, C. Deffayet and D. Langlois, hep-th/9905012.

\bibitem{vilipsi} A. Vilenkin, \PL {\bf 133} 177 (1983);
J. Ipser and P. Sikivie, \PR {\bf D30} 712 (1984).

\bibitem{nohair} G.W. Gibbons and S.W. Hawking,
{\it Phys. Rev.} {\bf D15}, 2738 (1977); 
S.W. Hawking and I.G. Moss,
{\it Phys. Lett.} {\bf B110}, 35 (1982); 
W. Boucher and G.W. Gibbons,
in {\it The Very Early Universe}, ed. G.W. Gibbons,
S.W. Hawking and S. Siklos, Cambridge University Press,
Cambridge (1983); A.A. Starobinsky, {\it JETP Lett.}
{\bf 37}, 66 (1983); 
R. Wald, {\it Phys. Rev.} {\bf D28}, 2118 (1982).

\bibitem{lyth} D. Lyth, hep-ph/9810320.

\bibitem{edi}
E. Halyo, hep-ph/9904432;
hep-ph/9905244.

\bibitem{andc} A. Linde, 
{\it Phys. Lett.} {\bf B129}, 177 (1983).

\bibitem{andh} A. Linde, {\it Phys. Lett.} {\bf B259}, 38
(1991); {\it Phys. Rev.} {\bf D49}, 748 (1994);

\bibitem{es}
E.J. Copeland, A.R. Liddle, D.H. Lyth, E.D. Stewart and D. Wands, 
\PR {\bf D49}, 6410 (1994);
E.D. Stewart,  \PR {\bf D51}, 6847 (1995).

\bibitem{Dterm} 
P. Binetruy and 
G. Dvali, {\it Phys. Lett.} {\bf B388},
241 (1996); E. Halyo, {\it Phys. Lett.} {\bf B387}, 43 (1996);
A.D. Linde and A. Riotto,
{\it Phys. Rev.} {\bf D56}, 1841 (1997).

\bibitem{banks}
T. Banks, M. Dine and A. Nelson, hep-th/9903019.

\bibitem{othercosmo}
H.A. Chamblin and H.S. Reall, hep-th/9903225;
P. Kanti and K.A. Olive, hep-ph/9903524;
J.M. Cline, hep-ph/9904495;
A. Riotto, hep-ph/9904485.

\bibitem{dvali}
G. Dvali and G. Gabadadze, hep-ph/9904221;
G. Dvali and M. Shifman, hep-th/9904021;
G. Dvali, hep-ph/9905204.

\bibitem{nih} T. Nihei, hep-ph/9905487.

\end{thebibliography}
\end{document}